\documentclass[english,preprint, aps]{revtex4-1}
\usepackage[LGR,T1]{fontenc}
\usepackage[latin9]{inputenc}
\usepackage{textcomp}
\usepackage{amsmath}
\usepackage{amssymb}
\usepackage{graphicx}

\makeatletter

\DeclareRobustCommand{\greektext}{%
  \fontencoding{LGR}\selectfont\def\encodingdefault{LGR}}
\DeclareRobustCommand{\textgreek}[1]{\leavevmode{\greektext #1}}
\ProvideTextCommand{\~}{LGR}[1]{\char126#1}

\providecommand{\tabularnewline}{\\}

\makeatother

\usepackage{babel}
\begin{document}

\title{Azimuthal Angular Decorrelation of Jets at Future High Energy Colliders}

\author{\.{I}. Ho\c{s}}
\email{ilknurhos@aydin.edu.tr}

\selectlanguage{english}%

\affiliation{Istanbul Ayd\i n University, Application and Research Center For
Advanced Studies, 34295, Istanbul, Turkey}

\author{H. Sayg\i n}
\email{hasansaygin@aydin.edu.tr}

\selectlanguage{english}%

\affiliation{Istanbul Ayd\i n University, Application and Research Center For
Advanced Studies, 34295, Istanbul, Turkey}

\author{S. Kuday}
\email{sinankuday@aydin.edu.tr}

\selectlanguage{english}%

\affiliation{Istanbul Ayd\i n University, Application and Research Center For
Advanced Studies, 34295, Istanbul, Turkey}

\begin{abstract}
The azimuthal angular decorrelation that is relevant to small-x QCD
physics is studied in this paper to show the BFKL effect with a recent
event generator. Events are generated at \textsurd s = 100 TeV with
proton-proton collisions and jets, that are reconstructed by the Anti-$k_{T}$
algorithm ($R=0.7$), with $p_{T}>35$GeV and in the rapidity range
of $|y|<6$ are selected for the study. The azimuthal-angle difference
between Mueller-Navelet Jets ($\Delta\text{\textgreek{F}}$ ) in the
rapidity seperation ($\Delta y$) up to $12$ is analysed. The distributions
of $<cosn(\pi-\Delta\Phi)>$ for $n=1,2,3$ and their ratio are also
presented as a function of $\Delta y$.
\end{abstract}
\maketitle

\section{INTRODUCTION}

The strong interaction between quarks and gluons, called partons,
are defined by the theory of Quantum Chromodynamics (QCD). According
to QCD, quarks carry colour charges (blue, red and green) and can
not be observed as free particles but in colourless states. This behaviour
is named as confinement. To study partons experimentally, one needs
to consider jets described by QCD in terms of pp scattering.

The momentum distribution functions of partons within the proton are
described by the evaluation equations, when running coupling as one
moves from one momentum scale to another. One of these evolution equations
is BFKL (Balitsky, Fadin, Kuraev, Lipatov)\citep{BFKL,BFKL-2,BFKL-3}
which describes the dependence on $x$, the parton momentum fraction.
BFKL equations require strong ordering of fractional momentum. An
ideal observable to study sensitivity of low-x QCD evolution and a
test of BFKL, is the distribution of azimuthal angle between two jets
in the same event. At the leading order calculations two jets are
back-to-back in xy-plane and perfectly correlated. However when energy
of collision increases the correlation of two jets break down by the
emission of an extra jet and the two selected jets are no longer back-to-back
in the xy-plane.

At Fermilab Tevatron, the studies of BFKL effect are performed at
\textsurd s = 1.8 TeV, 1800 and 630 GeV by D0 experiment \citep{D0-1,D0-2}.
At both studies, $\text{\textgreek{D}\textgreek{h}}$ is selected
up to 6, which could limit the observation of decorrelation effect.

The experiments at LHC provide larger rapidity seperation and higher
center-of-mass energy for the studies. The publication from ATLAS
Collaboration \citep{atlas-bfkl} is performed at \textsurd s = 7
TeV using jets with $p_{T}>20GeV$ and rapidity seperation of $|\text{\textgreek{D}}y|<6.$
and concludes that PYTHIA\citep{pythia8} and HERWIG\citep{herwig}
give best description of data as function of mean transverse momentum,
$\bar{p}_{T}$, while PYTHIA gives the best description of data as
a function of $\Delta y$. CMS publications \citep{CMS-kFactor,CMS-AzDecPaper}
report the measurements at 7 TeV and with jets $p_{T}>35GeV$ and
in the rapidity region of $|y|<4.7$. These publications indicate
that one can observe the BFKL effect at higher center-of-mass energies.

Recently, new projects based on higher collision energies are considering
to extend the search being performed in LHC. As one of these projects,
Future Circular Collider (FCC) \citep{FCC-1,FCC-2} is planning to
be built on 80-100 km tunnel to reach the 100 TeV collision energy.
FCC is hosted by CERN and planned to develop three accelerator facilities;
FCC-hh, FCC-he and FCC-ee. Another future collider project is Circular
Electron Positron Collider (CEPC) with the design of Super Proton-Proton
Collider (SPPC) \citep{CEPC-1,CEPC-2}. The CEPC-SPPC is planned to
have baseline of 100km circumference and to reach the center-of-mass
energy of 100 TeV. Since there is no available data at 100 TeV, to
see the BFKL effect at such center-of-mass energies one can simulate
hard QCD events and use jet reconstruction algorithms to investigate
kinematic distributions. 

\section{SMALL-x PHYSICS}

According to Quark-Parton Model, protons are made of point-like particles,
called partons. Thus when two hadrons collide, quarks and gluons inside
incoming hadrons interact with each other, indeed. The decrease of
the dependence of the structure functions on energy scale $Q^{2}$
is predicted by increasing center-of-mass energies. Later, structure
functions become the function of $x$ alone. The probability of finding
a parton carrying a fraction $x$ of the momentum of the proton is
defined by parton distribution functions (PDFs). The change in parton
distributions with variation in momentum scale is described by evolution
equations, which are calculated as perturbatively. The BFKL evolution
equation becomes valid at small-$x$ values. BFKL allows the resummation
of terms with $\left(\alpha_{s}log\left(\frac{1}{x}\right)\right)^{n}$
(where $\alpha_{s}$ is strong coupling) so it is valid at $log\left(\frac{Q^{2}}{Q_{0}^{2}}\right)\text{\ensuremath{\ll}}log\left(\frac{1}{x}\right)$.

Jets are particle sprays emitted from hadron-hadron collisions. The
measurement of jet shapes allows one to study the transition between
a parton, coming from hard scattering, and a hadron, observed experimentally.
Mueller Navelet jets (MN Jets) \citep{MNJets} are produced during
hadron-hadron collisions at high energies. These jets carry the longitudinal
momentum fraction of their parent hadrons. Thus, each jet is closed
to its parent hadron and that cause a large rapidity seperation between
jets. Such process allows to study the BFKL evidence. During the collision,
dijets are correlated and can be observed back-to-back or with different
angles. Back-to-back jets are balanced in transfer momentum and perfectly
correlated. But when an extra jet is emitted in such process, the
decorrelation begins to appear and break down of back-to-back topology
increase.

There are two main observables can show the effect of BFKL. One is
azimuthal angle between MN Jets ($\Delta\text{\textgreek{F}}$) with
respect to rapidity seperation between jets as suggested at \citep{DelDuca}.
If jets are highly correlated and back-to-back there should be a sharp
peak in the distribution of $\Delta\text{\textgreek{F}}$. However
by increasing rapidity seperation between jets, the $\Delta\text{\textgreek{F}}$
peak decrease and distribution becomes wider. The second observable
is average cosine value of $\Delta\text{\textgreek{F}}$ ($<cos(\pi-\Delta\Phi)>$)
\citep{Kim-1996,D0-1}. When jets are back-to-back and perfectly correlated,
one should expect a flat distribution at 1 in the plot of $<cos(\pi-\Delta\Phi)>$.
When decorrelation arise with extra jet emission, $<cos(\pi-\Delta\Phi)>$
value start to decrease with increasing rapidity seperation. And third
observable can be ratio of $<cosn(\pi-\Delta\Phi)>$ for different
$n$ values. That plot can show a clear change of $<cos(\pi-\Delta\Phi)>$.

\section{EVENT AND JET SELECTION}

Pythia8 \citep{pythia8} is used to generate the hard QCD events with
proton-proton collisions at \textsurd s = 100 TeV. $30\times10^{6}$
events are generated and during the generation of events, FastJet\citep{fastjet}
is used to reconstruct the jets by Anti-$k_{T}$\citep{anti-kT} jet
algorithm with cone radius of 0.7. As the preselection criteria, events
with at least two jets are used and jets are required to pass $p_{T}$
cut of 10 GeV and to be in the rapidity region of $|y|<7.$ 

In the analysis, following criteria are applied to select the jets:
\begin{itemize}
\item $p_{T}$ higher than 35 GeV 
\item in the rapidity region of $|y|<6.$ 
\item apply rapidity ordering of jets for each event and choose the jets
with highest rapidity and lowest rapity value, and name them most
forward jet and most backward jet, respectively. 
\end{itemize}
By that selection, in each event these two jets would have largest
rapidity seperation and can be named as Mueller-Navelet jets. The
control plots are produced to see the effect of selection criteria.
Figure 1 shows $p_{T}$ distributions of forward and backward jets
of which have $p_{T}$ higher than 35 GeV. It is clear that during
the analysis both most forward and most backward jets have $p_{T}>35GeV$
are used. In Figure 2, rapidity of MNJets is plotted, while phi distribution
is shown in Figure 3. These two figures show most of the jets are
in back-to-back in $xy$ plane. 

The number of events and number of jets before and after cuts are
presented in Table 1. Table 2 shows the number of jets at each $\Delta\text{\textgreek{F}}$
distribution at \textsurd s = 100 TeV. The number of jets decrease
with large rapidity seperation.

\begin{table}
\caption{Number of events and number of jets before and after cuts at \textsurd s
= 100 TeV}

\begin{tabular}{|c|c|c|c|}
\hline 
@\textsurd s = 100 TeV & Before Cuts & Events with at least 2 jets \& jet pT > 35 GeV & After MNJets Selection Criteria\tabularnewline
\hline 
\hline 
Number of Events & 3e+07 & 4.3733e+06 & 3442\tabularnewline
\hline 
Number of Jets & 1.92805e+12 & 3.28932e+10 & 1.41547e+07\tabularnewline
\hline 
\end{tabular}
\end{table}

\begin{table}
\caption{Number of jets at each $\Delta\text{\textgreek{F}}$ distribution
at \textsurd s = 100 TeV}

\begin{tabular}{|c|c|c|c|c|}
\hline 
 & $|\text{\textgreek{D}}y|<3.$ & $3.<|\text{\textgreek{D}}y|<6.$ & $6.<|\text{\textgreek{D}}y|<9.$  &  $9.<|\text{\textgreek{D}}y|<12.$ \tabularnewline
\hline 
\hline 
Number of Jets & 3427680 & 4465153 & 3353053 & 2904755\tabularnewline
\hline 
\end{tabular}
\end{table}

\begin{figure}
\includegraphics[scale=0.5]{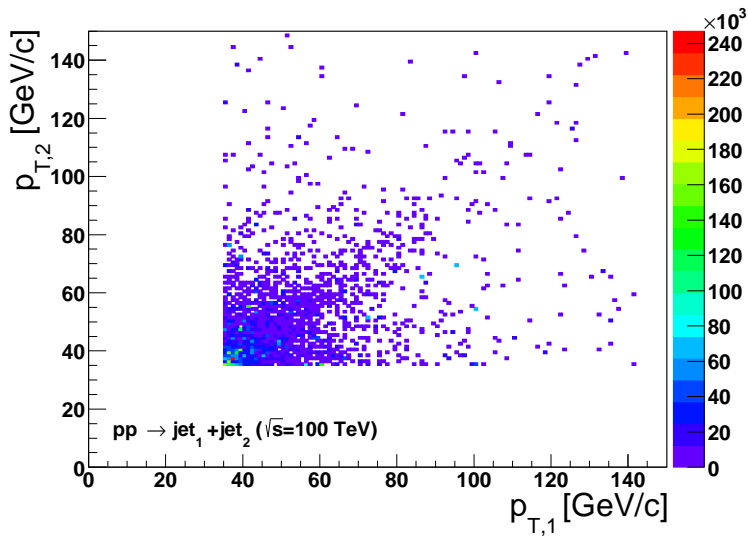}

\caption{Forward jet $p_{T}$ vs backward jet $p_{T}$ }
\end{figure}

\begin{figure}
\includegraphics[scale=0.5]{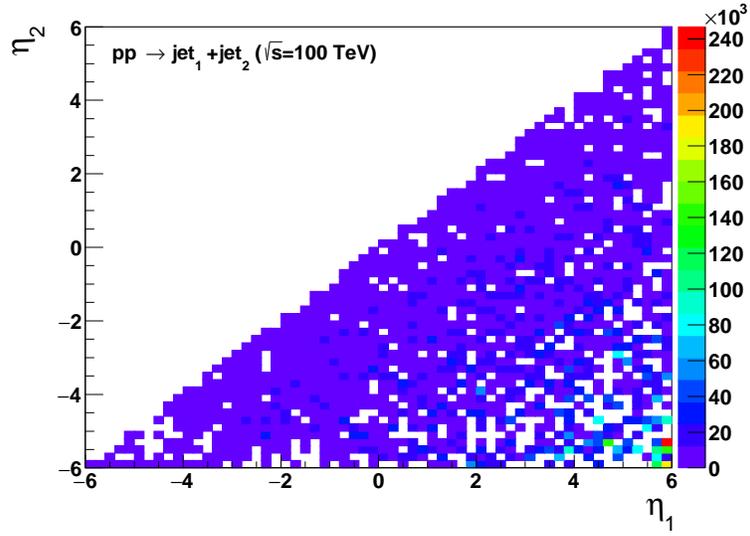}

\caption{Forward jet rapidity vs backward jet rapidity }
\end{figure}

\begin{figure}
\includegraphics[scale=0.5]{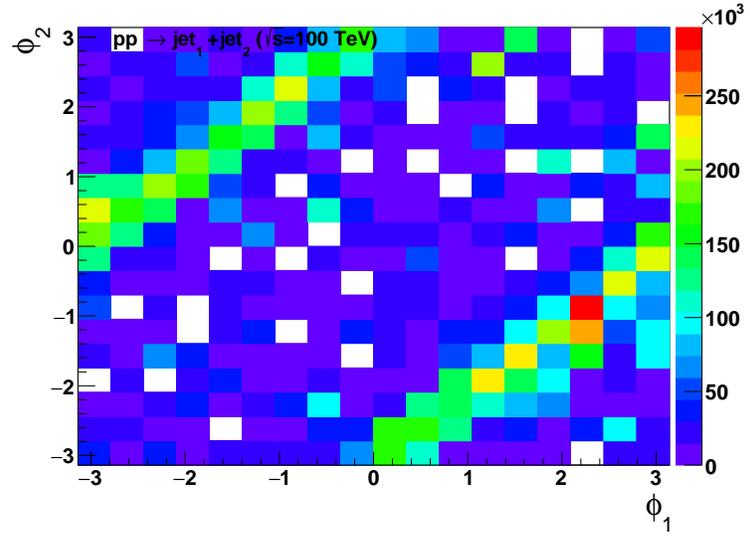}

\caption{Forward jet phi vs backward jet phi}
\end{figure}

\section{ANALYSIS}

After being sure about selection of MN Jets, the result plots are
produced. The first observable to see the signs of BFKL effect, the
azimuthal-angle difference between MNjets ($\Delta\text{\textgreek{F}}$
) as a function of the rapidity seperation of MNJets, is shown in
Figure 4. As described in previous section, the decorrelation will
rise by increasing rapidity seperation between jets. To see the effect
clear, the distribution is plotted for four rapidity seperations:
$|\text{\textgreek{D}}y|<3.$, $3.<|\text{\textgreek{D}}y|<6.$, $6.<|\text{\textgreek{D}}y|<9.$,
and $9.<|\text{\textgreek{D}}y|<12.$ If one check first binning of
the histogram, then can see clearly $|\text{\textgreek{D}}y|<3.$
is in the top while in the last binnings it is in the bottom and $9.<|\text{\textgreek{D}}y|<12.$
appears in the top. That shows with increasing rapidity between jets,
the peak of $\Delta\text{\textgreek{F}}$ distribution decrease and
the distribution becomes wider comparing to the distributions with
narrower $\Delta y$.

The second observable to see the BFKL effect, $<cos(\pi-\Delta\Phi)>$
distribution between MN Jets, is shown in Figure 5. $<cosn(\pi-\Delta\Phi)>$
is presented for $n=1,2$ and $3$ with different line colors. The
distribution start around 1 and then decreases for larger rapidity
seperation. When $n$ increase, the change in the distribution becomes
more significant. There are some statistical fluctuations in the distributions
are observed.

Figure 6 shows the ratio of $<cos2(\pi-\Delta\Phi)>$ to $<cos(\pi-\Delta\Phi)>$
($\frac{C_{2}}{C_{1}}$ , left plot) and $<cos3(\pi-\Delta\Phi)>$
to $<cos2(\pi-\Delta\Phi)>$ ($\frac{C_{3}}{C_{2}}$ , right plot)
as a function of the rapidity separation $\Delta y$. Except the bins
with low statistics, the general behaviour of distribution is decreasing
as a function of $\Delta y$.

\begin{figure}
\includegraphics[scale=0.5]{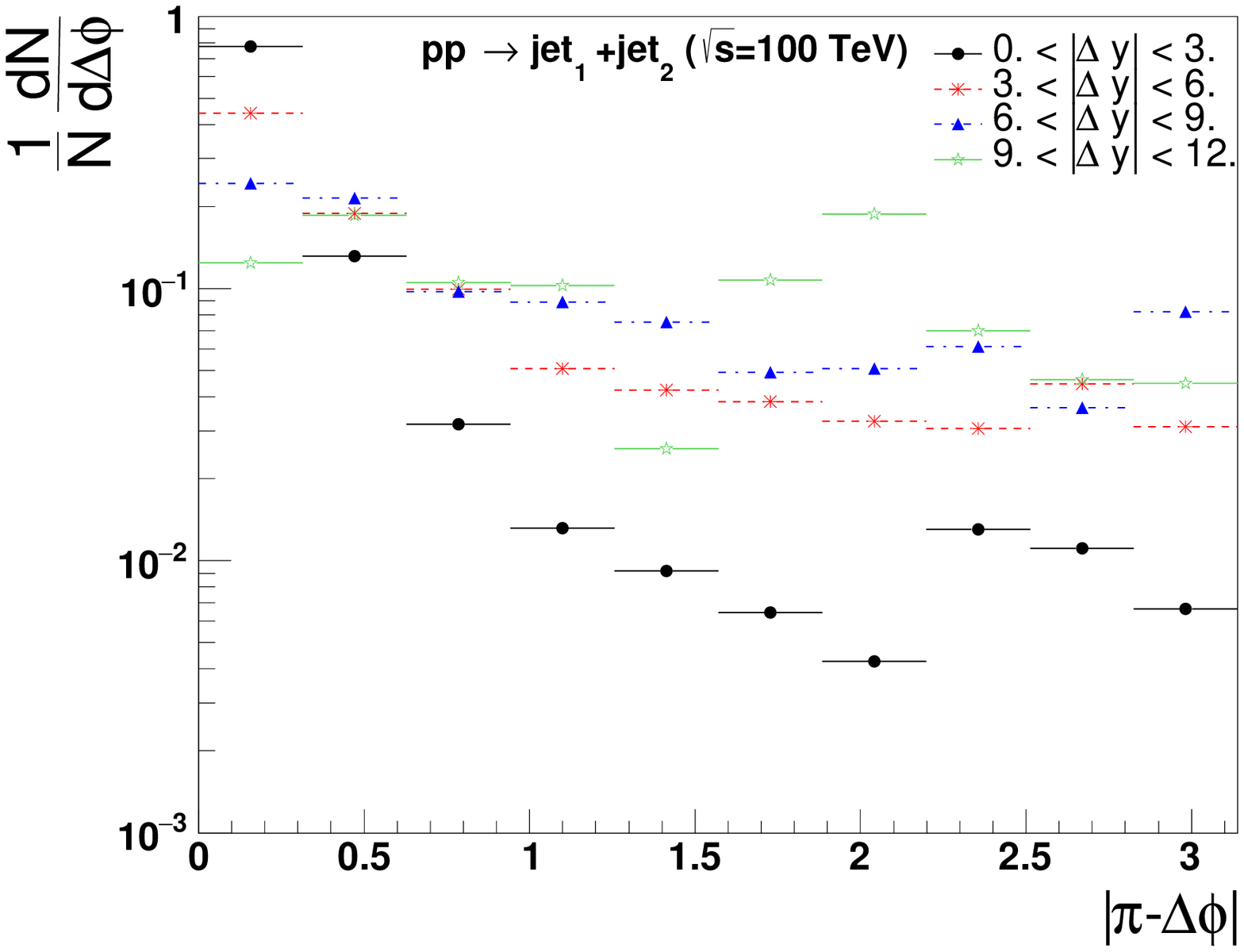}

\caption{The azimuthal-angle difference between MNjets ($\Delta\text{\textgreek{F}}$
) in the rapidity of $|\text{\textgreek{D}}y|<3.$, $3.<|\text{\textgreek{D}}y|<6.$,
$6.<|\text{\textgreek{D}}y|<9.$ and $9.<|\text{\textgreek{D}}y|<12.$
at \textsurd s = 100 TeV}
\end{figure}

\begin{figure}
\includegraphics[scale=0.5]{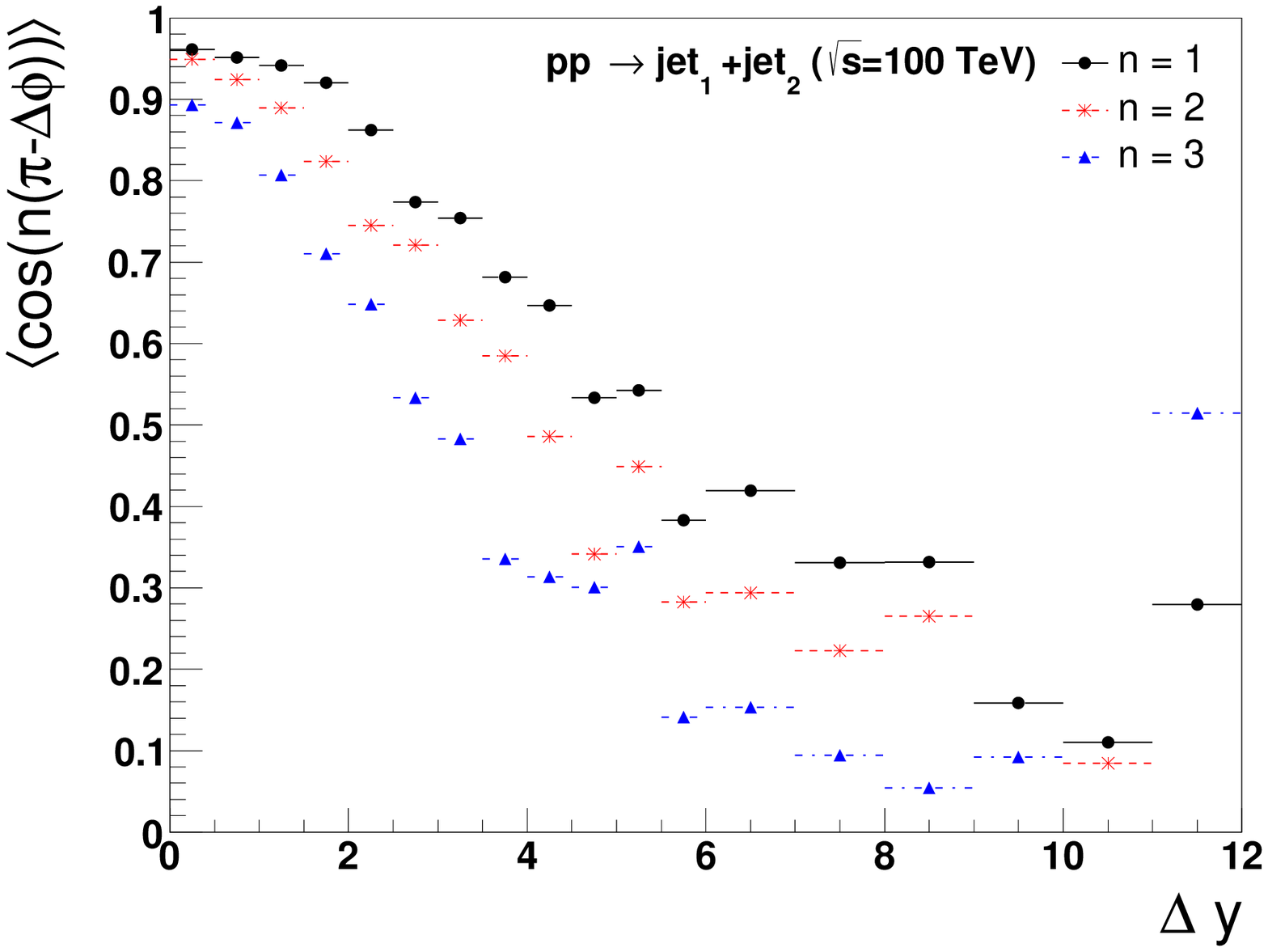}

\caption{$<cos(\pi-\Delta\Phi)>$, $<cos2(\pi-\Delta\Phi)>$ and $<cos3(\pi-\Delta\Phi)>$
as a function of $\Delta y$ at \textsurd s = 100 TeV}
\end{figure}

\begin{figure}
\includegraphics[scale=0.4]{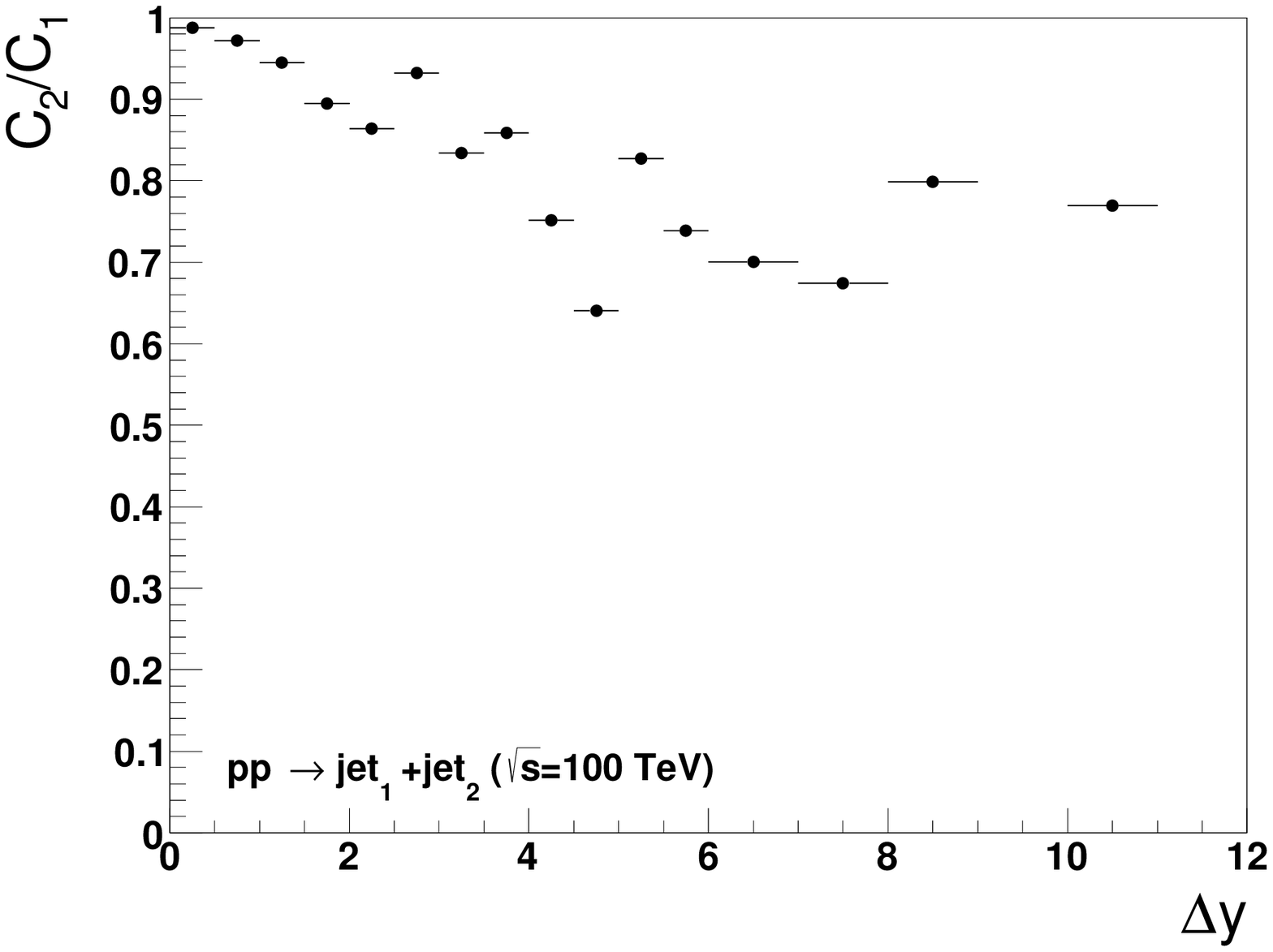}\includegraphics[scale=0.4]{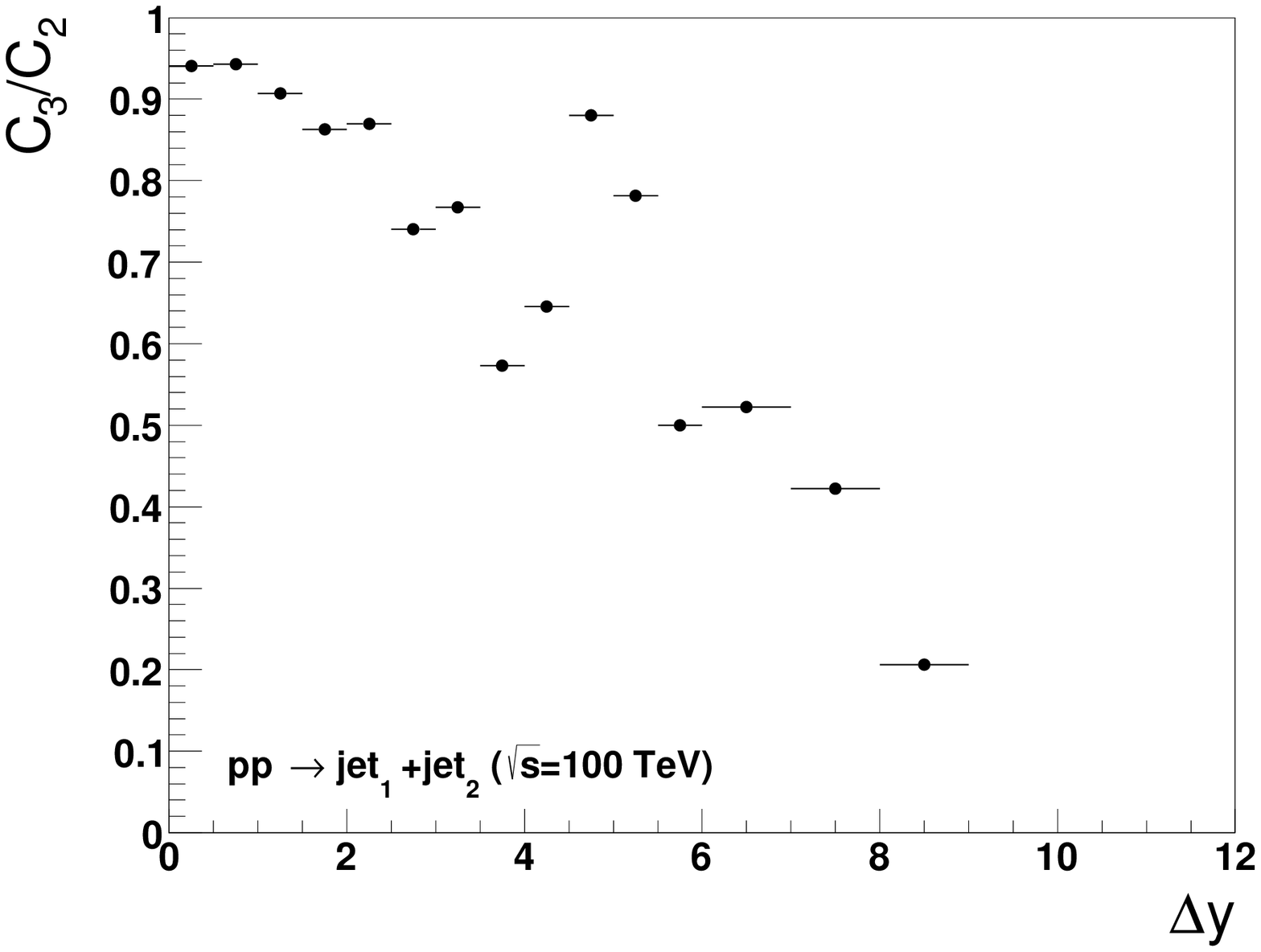}

\caption{Ratio of average cosine $\frac{C_{2}}{C_{1}}$ (left) and $\frac{C_{3}}{C_{2}}$
(right) as a function of $\Delta y$ at \textsurd s = 100 TeV}
\end{figure}

\section{CONCLUSION}

In the analysis, we have showed a significant difference between the
kinematic distributions of MNjets that are selected to have largest
rapidity seperation with $p_{T}>35GeV$ at \textsurd s =100 TeV using
Pythia8 event generator and FastJet clustering algorithm which are
considered as one of the most realistic and updated tools in data
analysis. In Figure 4, one can see that the peak of azimuthal-angle
distributions of MNJets are decreasing with selected rapidity regions
($\Delta y$) up to 12 and getting wider by the function of $\Delta y$.
We have found that the distributions of jets for rapidity regions
of $|\text{\textgreek{D}}y|<3.$, $3.<|\text{\textgreek{D}}y|<6.$,
$6.<|\text{\textgreek{D}}y|<9.$, $9.<|\text{\textgreek{D}}y|<12.$
are 24.2\%, 31.6\%, 23.7\% and 20.6\%, respectively. The average cosine
value of $\Delta\text{\textgreek{F}}$ is also decreases with the
increasing $\Delta y$. In figure 5, same effect can be seen in the
ratio plots of $<cosn(\pi-\Delta\Phi)>$ for $n=1,2,3$ except the
bins suffering from low statistic.

In conclusion with increasing center-of-mass energies, the observables
of BFKL effects become significant for jet based analysis particularly
in forward regions of detectors and can only be justified by collected
data at future high energy colliders such as FCC and CEPC-SPPC. Therefore
one should consider the BFKL evolution equation and parton momentum
fraction ($x$) dependency in the parton distribution functions in
the analysis of experimental data in order to achieve more accurate
results from jet based analyses.

\end{document}